# Research on material identification of mobile phones falling to the ground

Xuesong Wang[&,1], Xiaoyu Lei[&,1], Yi Zhou[2], Dongsheng Zhang[*,1]
(&:These two authors contributed equally to this work and should be considered co-first authors.
[1]Xi'an Jiaotong University, Xi'an 710064 China
[2]Beijing Institute of Space Launch Technology, Beijing 100000 China)

**Abstract**: The failure mode of the phone falling has a lot to do with the ground material. At present, the research on ground material and mobile phone damage is generally carried out through experiments, which is extremely costly. This paper presents a method to identify the material of mobile phones falling on the ground. The method determines the material of the mobile phone falling to the ground according to the data of the mobile phone accelerometer and can obtain the ground material of the mobile phone falling through a large number of user data. By analyzing the physical process of mobile phone falling, the accelerometer data interval which can reflect the characteristics of falling is reasonably intercepted. And the data features that can reflect the collision are extracted. Finally, based on the fully connected neural network, the method of determining the material of mobile phones falling on the ground is developed. The experimental results show that the method has a high identification rate, and the average identification rate for different ground materials is 96.75%. Instead of relying on auxiliary devices, the identification process only relies on the phone's built-in acceleration sensor. This paper fills the gap in the field of mobile phone falling ground material identification. It has a great contribution to the collection of user drop data, the analysis of the drop process, and the improvement of mobile phone design defects.

## 1   Introduction

When the mobile phone falls and hits the ground, its ground material has an important effect on the failure mode of the mobile phone. Researching the damage incurred by mobile phones when dropped on different ground materials contributes to continuously optimizing mobile phone structures, thereby enhancing their impact resistance capabilities [1–7]. Currently, investigating the damage sustained by mobile phones when dropped on different ground materials relies heavily on conducting numerous drop tests. This method incurs significant costs and requires a substantial number of test devices. Therefore, there is an urgent need to develop a low-cost and effective method for identifying the ground materials encountered during mobile phone falls.

A promising approach involves collecting data from users' mobile phones via the cloud, enabling the acquisition of extensive information on the damage caused by falls onto various ground materials without the need for numerous physical drop tests. This method provides valuable guidance for the structural optimization of next-generation mobile phones and significantly reduces research and development costs. Additionally, this approach offers the advantage of statistically identifying the types of ground materials on which mobile phones are most frequently dropped, thereby allowing manufacturers to tailor their structural enhancements accordingly. However, the challenge lies in accurately extracting information about the ground material from the data of damaged mobile phones.

Current methods for identifying ground materials primarily rely on sensors such as visual and tactile sensors to detect the characteristics of the ground. The collected data from these sensors are processed to accurately identify the type of ground material. Matti et al. [8] proposed a material classification method based on the acceleration, sound waves, and image information generated when tapping the surface of an object. The information on 69 texture surfaces was recorded manually, and the classification accuracy reached 74%. Zheng et al. [9] identified the material of the object according to the acceleration signal generated when the hand-held rigid object scratched the surface, combined with the image information of the surface texture. Experiments show that the proposed method is robust and effective. Wei et al. [10] proposed a model to identify the auditory and tactile data (acceleration, normal force, and friction) generated by the tool dragged along the surface of an object. The maximum classification accuracy is 87.55%. Jamali et al. [11] identified ground materials through the surface problem signals perceived by bionic fingers. Zhu et al. [12] improved the vision-based ground material identification method. The highest identification accuracy was 96.56%.

These kinds of ground material identification methods[13,14] above are generally based on the tactile and auditory information of the tool crossing the surface of the object. The data characteristics collected by these methods are distinct from those observed during mobile phones' impact on the ground. When a mobile phone falls and makes contact with the ground, it generates a momentous impact, characterized by an extremely brief duration. Other methods use the visual information of the surface texture. Although mobile phones are equipped with cameras, the sudden and brief nature of falls makes it challenging to promptly identify the falling state of the device and activate the camera application to capture images of the ground. Therefore, the above method is difficult to apply directly.

To facilitate software upgrades for user smartphones via Over-The-Air (OTA) updates and acquire fall impact data from existing user devices via the cloud, it's necessary to utilize the signals from sensors embedded within the smartphones. These sensors typically include gyroscopes, gravity sensors, and proximity sensors. By analyzing the principles and characteristics of these sensors [15–19], this study asserts that the accelerometer is better suited to reflect the characteristics of the mobile phone's fall and collision process. It is an appropriate source of analytical data for determining the ground material during smartphone falls.

This paper addresses the gap in the field of identifying ground materials in mobile phone falls by proposing a method for recognizing the ground material during mobile phone falls. It opens up the possibility of acquiring data on mobile phones falling on different ground materials via the cloud. By precisely cutting accelerometer data based on a physical model of the falling process, extracting features from the cut data, and employing a neural network for classification based on the extracted features, this method ultimately identifies the ground material of mobile phone falls. The proposed approach not only saves costs associated with drop tests, significantly reducing development expenses but also enables manufacturers to optimize mobile phone structures by statistically identifying the types of ground materials on which mobile phones are most frequently dropped.

The specific structure of this paper is as follows: Chapter 2 introduces the method of cutting data from accelerometer data to capture the falling process. Chapter 3 discusses the analysis and processing of accelerometer data during the falling process to achieve the identification of the ground material of the fall. This includes processes such as feature extraction and neural network classification. Chapter 4 presents experimental validation. Different experiments are conducted with varying drop heights and initial poses on different ground materials to acquire accelerometer data, confirming the effectiveness of the proposed ground material identification method. Chapter 5 provides a summary of the entire work.

# 2 Data cutting of mobile phone fall and touchdown process based on human cognitive model

The accelerometer data collected during the experiment includes several processes. For example, the mobile phone is held in the hand, the mobile phone is stably placed on the ground and the mobile phone touches the ground. Among them, only when the phone touches the ground, the phone is in contact with the ground material. Only in this state can the accelerometer data of the mobile phone reflect the characteristics of the ground material to a certain extent. Therefore, to identify the ground material when the mobile phone falls and touches the ground based on the accelerometer data, it is necessary to cut the accelerometer data.

Figure 1 shows the visual curve of the square root of the accelerometer during the drop of the mobile phone. Based on human cognition of the principle of the mobile phone accelerometer and the force of the mobile phone during the whole fall process, the curve can be divided into four regions:

① Stable area before falling. At this time, the phone is held in the hand and the value of the acceleration jitter around the value of the acceleration of gravity.

② Weightless area. During this time, the phone slipped from the hand and entered a state of weightlessness. The value of the accelerometer is significantly lower than normal and lasts for some time. The length of time duration is related to the height of the phone from the ground when it falls. Although the value of the accelerometer around 3s in the figure is also low, it does not belong to the weightlessness region of the fall. It is because the phone is weightless after the collision bounces, resulting in a lower magnitude of the accelerometer.

③ Mobile phone contact impact area. This region follows the weightless area, and the accelerometer values continuously undergo significant mutations. The moment the accelerometer's value changes for the first time is the moment of touch $t_c$.

④ Mobile phone stability area. The phone is stationary on the ground, and the value of the accelerometer in the area is stable around the local gravitational acceleration value, with only slight jitter. The starting point of this region is the end of the fall process $t_w$.

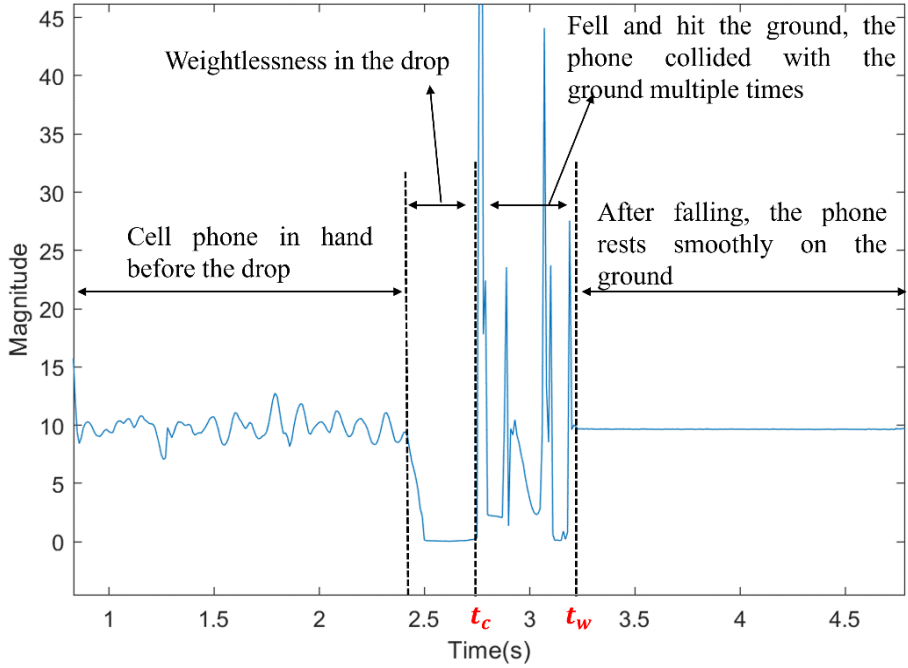

Figure 1 Human cognition of accelerometer data during the fall of mobile phone

When humans observe and recognize this curve, the time $t_c$ when the mobile phone touches the ground and the time $t_w$ when the fall process ends can be easily identified. Then the data between $t_c$ and $t_w$ can be intercepted for the input data of the ground material identification algorithm. In this paper, the process interpretation of human cognitive observation of the visual curve of the accelerometer mode and the location of the touchdown time $t_c$ and the end time $t_w$ of the fall process is converted into a program algorithm. $t_c$ and $t_w$ are automatically detected and the data between $t_c$ and $t_w$ is intercepted. Then it can be input into the ground material identification algorithm for processing. The basic flow of the data-cutting algorithm is as follows.

1. Locate the weightless area of the fall

In the process of identifying the beginning time of the fall and the moment of touching the ground, how to accurately locate the weightless area of the fall is particularly important. The sliding window method is adopted here. The root-mean-square of the data in the window of the accelerometer value of the whole falling process of the mobile phone is computed. The calculated results represent the energy in the window, and the earliest low-energy region is found, that is, the area of falling weightlessness. This step corresponds to the human perception of the drop zero gravity process on the curve of the accelerometer value. The value of the accelerometer is significantly lower than the value of the acceleration of gravity for a short period.

2. Determine when the phone hits the ground

In the data-cutting algorithm described in this article, it is necessary to determine the moment when the phone falls and first touches the ground. This moment immediately follows the drop to the weightlessness zone, and the accelerometer value changes significantly. According to this feature, the time of touchdown $t_c$ can be located by identifying the time of the first significant change of the accelerometer value after the weightlessness region.

3. Determine when the fall process ends

After the fall to the ground, the phone is stationary on the ground. The accelerometer values in this region are stable around the local gravitational acceleration values, with only slight jitter. The beginning time of this region is the end time of the falling process $t_w$. If the value of acceleration at a certain time and its subsequent consecutive 0.2s is stable near the local gravitational acceleration value, and the amplitude jitter is less than the threshold value, it is identified that the moment is the end of the drop process $t_w$.

4. Data cutting

The pseudo-code for the data-cutting algorithm is shown below.

---

Algorithm: Data-cutting

---

Input: Cell phone accelerometer data collected during a drop in the experiment *acc_m*
Output: Mobile phone accelerometer data after cutting, including only the mobile phone touch collision process *acc_m_cut*

% Parameter initialization
% Window size
*windo_size*
% Sliding step size
*m_step*

```
% power
power_
% Threshold value
F_c
F_w
% Local gravitational acceleration value
G_d

% Calculate the whole process power value
jj = 1;
for i =1:m_step:(length(acc_m)-windo_size-1)
    power_(jj) = sum(acc_m(i:i+windo_size-1).^2)/windo_size;
    jj = jj + 1;
end

% The moment of minimum power
min_power = min(power_);
stand_power = G_d ^2;
[min_pow, p_] = min(power_);

% Moment of touchdown
for ii = p_+windo_size : length(acc_m)
    if acc_m(ii) > 9.8*F_c
        t_c = ii;
        break
    end
end

% The moment the fall ends
for k = t_c : length(acc_m)
    count_k = 0;
    for m = 1 : 20
        if   G_d - F_w < acc_m(k+m) < G_d+ F_w
            count_k = count_k + 1;
        end
    end
    if count_k =20
        t_w = k;
```

```
        end
end

% Data cutting
acc_m_cut = acc_m(t_c:t_w);
```

Figure 2 is a comparison diagram of accelerometer data before and after cutting by the above algorithm. It can be seen that the cutting results are in line with human cognition and physical laws. After more than 100 times of data cutting experiments, the results of data cutting are compared with human cognition, and the coincidence degree is more than 95%.

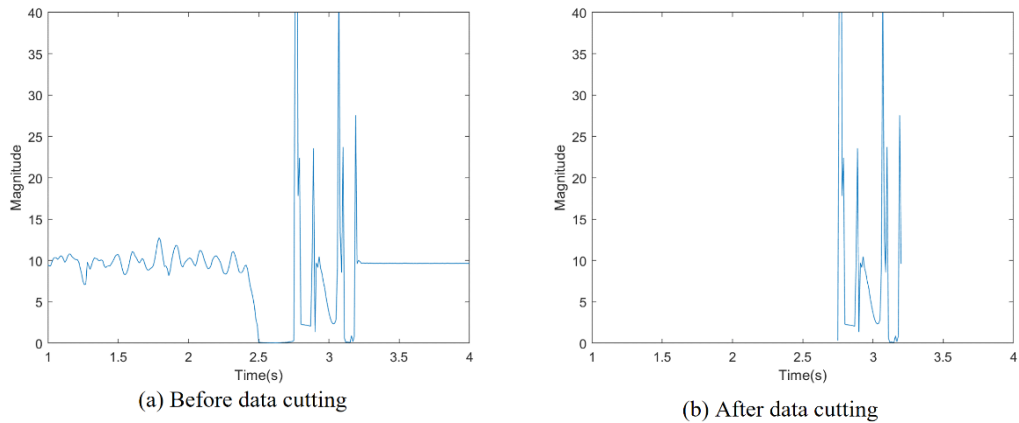

(a) Before data cutting     (b) After data cutting

Figure 2 Sketch of accelerometer data before and after data cutting

## 3   Ground material identification method

After intercepting accelerometer data related to ground features. In this paper, feature extraction of data is carried out first. Then the deep learning method is used to identify the ground material. According to the time domain information of the drop data, the neural network is constructed. The identification of five different floor materials including quilt, carpet, asphalt, granite, and marble is realized.

### 3.1 Data feature extraction

Because the length of the intercepted data is different, it is difficult to input it directly into the neural network for training. Therefore, feature extraction is performed on each piece of data, and after processing, the data feature dimensions corresponding to each drop process are consistent. It contains some basic time domain features, as well as features found in the study that are valuable for ground material identification. The data characteristics and meanings used by the algorithm in this paper are shown in Table 1. Where $accm$ is the clipped accelerometer data and $n$ is the data length.

Table 1 Mobile phone drop related data feature

| Feature | Formula or interpretation |
| --- | --- |
| Maximum value  $MAX$ | $\max(accm)$ |
| Minimum value  $MIN$ | $\min(accm)$ |
| Peak-to-peak value  $PPV$ | $MAX - MIN$ |
| Mean value  $MEAN$ | $\frac{1}{n}\sum_{i=1}^{n} accm(i)$ |
| Root amplitude  $RV$ | $\frac{1}{n}\sum_{i=1}^{n}\sqrt{|accm(i)|}$ |
| Variance  $Var$ | $\frac{1}{n}\sum_{i=1}^{n}(accm(i)-MEAN)^2$ |
| Standard deviation  $SD$ | $\sqrt{Var}$ |
| Root-mean-square  $RMS$ | $\frac{1}{n}\sqrt{\sum_{i=1}^{n}accm(i)^2}$ |
| Warping  $SE$ | $\frac{1}{n}\sum_{i=1}^{n}(|accm(i)|-MEAN)^4$ |
| Skewness  $SK$ | $\frac{1}{n}\sum_{i=1}^{n}(|accm(i)|-MEAN)^3$ |
| Shape factor  $SF$ | $RMS / MEAN$ |
| Peak factor  $PF$ | $MAX / RMS$ |
| Pulse factor  $PulF$ | $MAX / MEAN$ |
| Margin factor  $MarF$ | $MAX / RV$ |
| Clearance factor  $ClF$ | $MAX / Var$ |
| Power  $power\_$ | $\sum_{i=1}^{n} accm(i)^2$ |
| Hit time  $t_c$ | The time it takes to hit the ground. |
| Peak time  $PeakM$ | The time it takes to reach the  $MAX$ . |
| $CountPV$ | The number of peaks and valleys. As shown |
| | in Figure 3. |
| *PWH* | At half the height of the peak-value difference, the width of the peak. As shown in Figure 3. |
| $PWHs_3$ | Sum of the *PWH* of the first three peaks. |
| *PWHs* | Sum of half *PWH* of all peaks. |
| *PWH'* | At the half of the absolute height of the peak, the width of the peak. As shown in Figure 3. |
| *PVn* | $CountPV \cdot n$ |
| *PVnp* | $CountPV \cdot n / PWH$ |

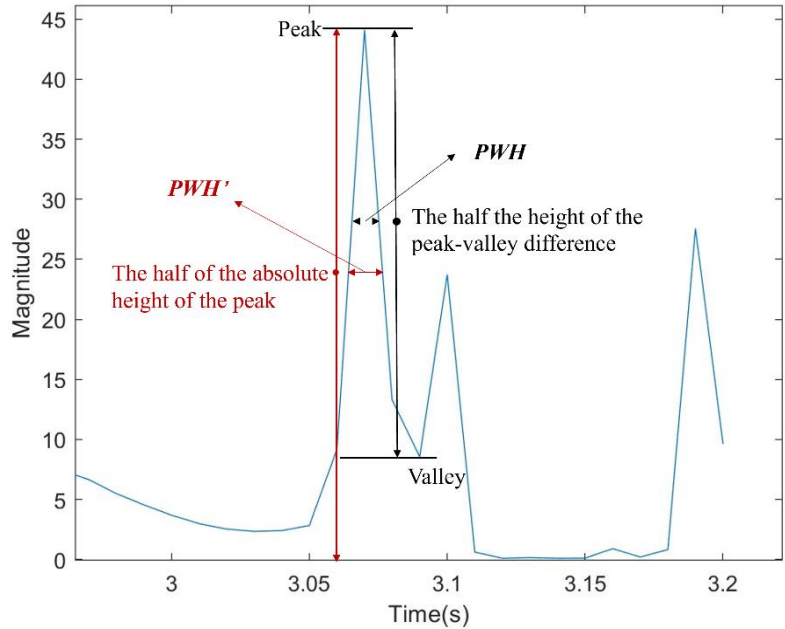

Figure 3 Diagram of *PWH* and *PWH'*

## 3.2 Ground material identification method based on neural network

As shown in Figure 4, this paper employs a fully connected neural network for ground material identification. It comprises two main components: forward propagation and backward propagation. During forward propagation, parameters are initialized, and layer outputs are computed. In contrast, during backward propagation, the gradient of the loss function is calculated to update weights and biases. In the backward propagation process, the output error is first computed through forward propagation, followed by gradient calculation layer by layer. Finally, the weights and biases are adjusted using gradient descent to establish the optimal mapping relationship between inputs and outputs.

The five-class ground material detection data set is divided into training set, verification set, and test set according to the ratio of 7:1:2. There are 3000 training sets, 200 verification sets, and 800 test sets. Add material label information to each time-domain feature information, that is, quilt, carpet, asphalt, granite, and marble materials are represented by labels 0-4.

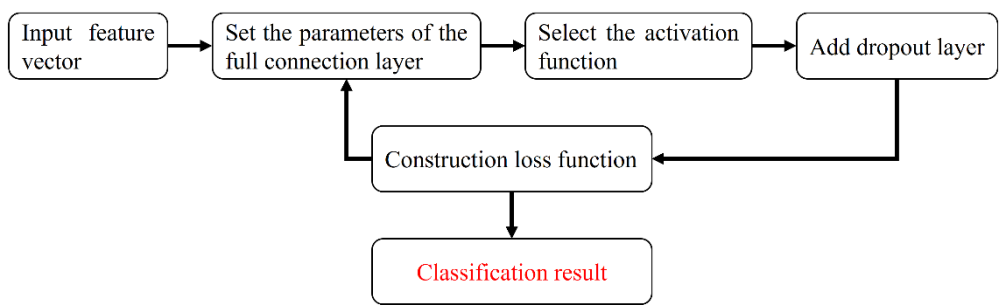

Figure 4 Ground material identification neural network diagram

# 4 Experimental verification

The ground materials to be identified in this paper are quilt, carpet, asphalt, granite, and marble. These ground materials are more common in life, and the degree of hardness is different, which is in line with the applicable scenario of mobile phone fall.

## 4.1 Data acquisition

In the experiment, accelerometer data during the mobile phone's fall and impact with the ground were collected using the accelerometer embedded in the mobile phone. To eliminate the influence of drop height and orientation on the identification results and to simulate user scenarios, manual experiments were conducted. In these experiments, a person held the phone and released it from various heights and poses. Five different ground materials were selected for the phone drop tests, including quilt, carpet, asphalt, granite, and marble, to simulate common scenarios of phone falls. The specific experimental conditions are detailed in Table 2.

During the drop test process, mobile phones can be damaged. Once they reach a certain level of damage, they become unsuitable for further drop tests, necessitating the replacement of the test device. The cost of each mobile phone is approximately $500. The research team allocated a significant amount of funds and ultimately conducted 4000 drop test experiments.

Regarding the experimental conditions outlined in Table 2, there were a total of 100 experimental conditions, with each condition repeated 40 times to ensure the reliability of the data.

Table 2 Phone drop test condition

| Height | Pose | Ground material |
|---|---|---|
| ① 0.4m | ①　The screen surface touches the ground. | ①　Quilt |
| ② 0.8m | ②　The back (camera side) touches the ground. | ②　Carpet |
| ③ 1.2m | ③　The long side touches the ground. (Figure 5(a)) | ③　Asphalt |
| ④ 1.6m | ④　The short side touches the ground. (Figure 5(b)) | ④　Granite |
|  | ⑤　The angle touches the ground. (Figure 5(b)) | ⑤　Marble |

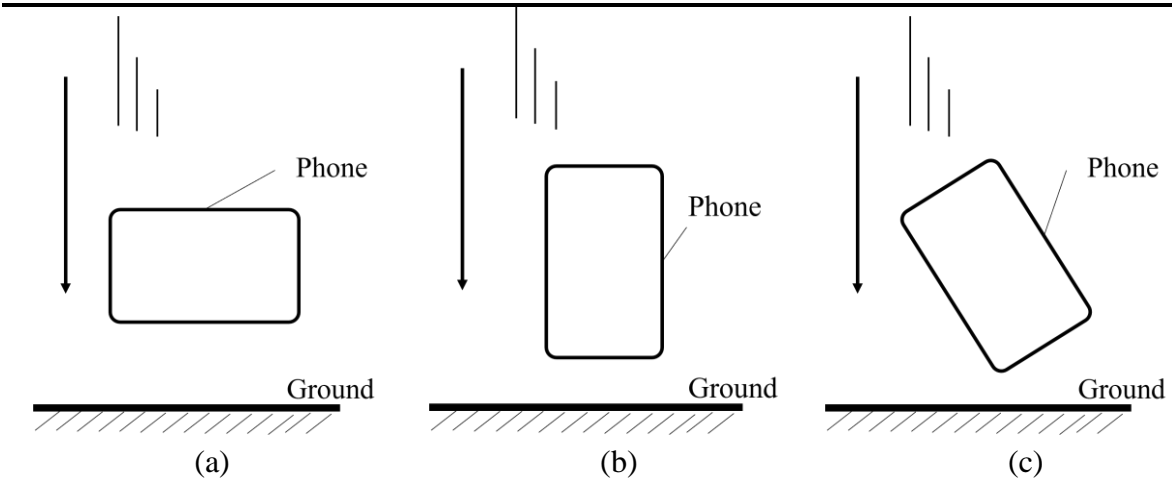

Figure 5 Diagram of different phone fall poses

　　The experiments described in this paper were conducted in the drop test laboratory of Honor Mobile Company. Due to the company's strict confidentiality regulations, no photographs of the experimental equipment or scenes can be taken out of the laboratory. Therefore, this paper cannot include images of the experimental setup or environment. The experimental data were provided to our research team with permission from Honor Company to carry out the surface material identification study. The accelerometer data of the smartphones falling on the five different surface materials are shown in Figure 6.

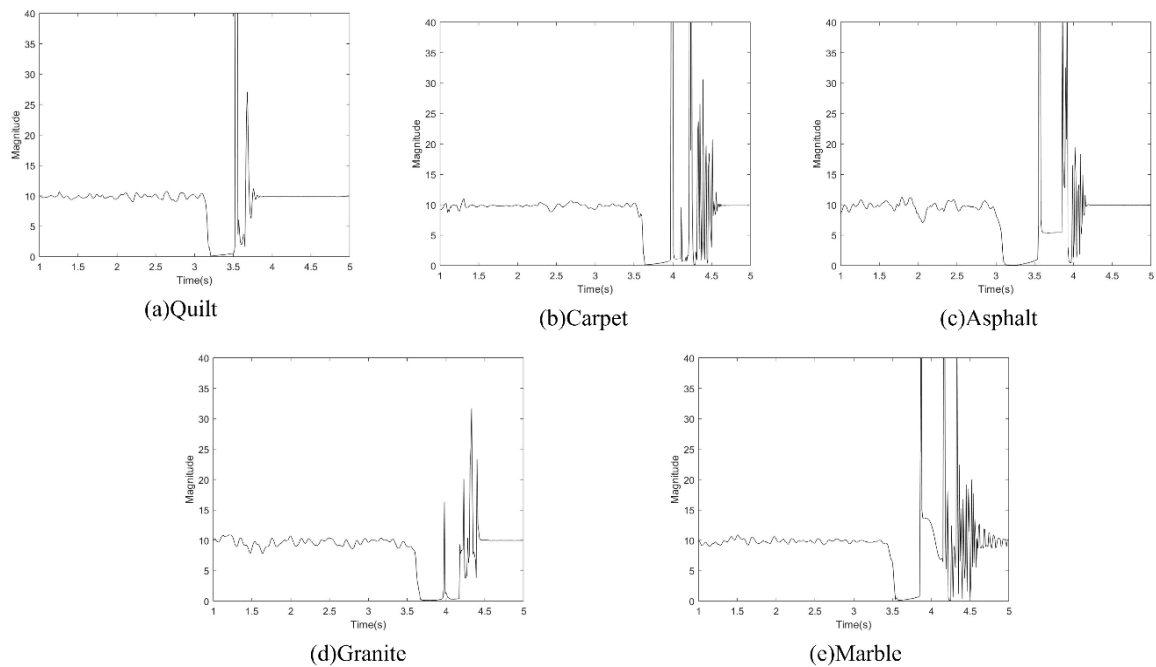

(a)Quilt　　　　　　　　　(b)Carpet　　　　　　　　　(c)Asphalt

(d)Granite　　　　　　　　　(e)Marble

Figure 6 Accelerometer magnitude of the phone falling on five different ground materials

## 4.2 Results

First of all, a large number of drop tests are carried out on different ground materials with the test prototype. The collected acceleration data are clipped and feature extracted. Then the neural network serves as the model for ground material classification. Utilizing the 25 features identified in Section 3.1, these features serve as inputs to the network. The training set and verification set of five different ground material samples obtained in the data preprocessing stage are input into the fully connected neural network. During training, the loss and accuracy of the verification set are recorded. Draw the line chart shown in Figure 7. The training time totaled 21.25 seconds.

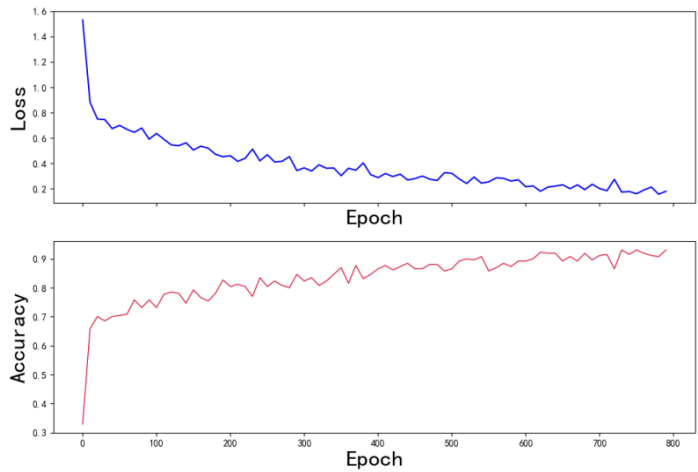

Figure 7 Line chart of training loss and accuracy in ground material detection

To verify the accuracy and robustness of the neural network in identifying ground materials. The trained model was used to make p identifications on 800 test sets. The identification time is 0.09 seconds, and the results are shown in Table 3. From the experimental results, it can be seen that the research on the ground material identification of mobile phone falling described in this paper can realize the identification of five common ground materials in life, including quilt, carpet, asphalt, granite, and marble. The accuracy of material identification is higher, reaching more than 95%, and the average accuracy is as high as 96.75%. This paper fills the gap in the field of mobile phone falling ground material identification.

Table 3 Ground material identification results statistical table

| Type | Result of identification | | | | | | Accuracy/% |
|---|---|---|---|---|---|---|---|
| | Quantity | Quilt 0 | Carpet 1 | Asphalt 2 | Granite 3 | Marble 4 | |
| Quilt 0 | 160 | 155 | 3 | 0 | 1 | 1 | 96.78 |
| Carpet 1 | 160 | 0 | 154 | 2 | 3 | 1 | 96.25 |
| Asphalt 2 | 160 | 2 | 1 | 154 | 1 | 2 | 96.25 |
| Granite 3 | 160 | 1 | 1 | 2 | 155 | 0 | 96.78 |
| Marble 4 | 160 | 0 | 2 | 1 | 1 | 156 | 97.50 |
| Total | 800 | 158 | 161 | 159 | 160 | 160 | 96.75 |

# 5  Conclusion

This paper proposes a method to identify the material of mobile phones falling on the ground. By analyzing the physical process of the mobile phone falling, the accelerometer data interval that can reflect the fall characteristics is reasonably intercepted, and the data features that can reflect the collision are extracted. Finally, based on the neural network, the method of determining the material of mobile phones falling on the ground is developed. The experimental results show that this method has a high identification rate, and the identification accuracy of different ground materials is more than 96%. And the whole process does not rely on auxiliary equipment, the identification process only relies on the built-in acceleration sensor of the mobile phone. This paper fills the gap in the field of mobile phone falling ground material identification. This method makes it possible to collect a large number of users' fall data and analyze the fall process. It has made outstanding contributions to improving the structural reliability of mobile phones and making up for design defects.